\documentstyle[12pt]{article}
\begin{document}
\newcommand{\be}{\begin{equation}}
\newcommand{\ee}{\end{equation}}
\newcommand{\bea}{\begin{eqnarray}}
\newcommand{\eea}{\end{eqnarray}}
\newcommand{\beas}{\begin{eqnarray*}}
\newcommand{\eeas}{\end{eqnarray*}}

\baselineskip 14 pt
\parskip 12 pt

\begin{titlepage}
\begin{flushright}
{\small hep-th/0403189}\\
\end{flushright}

\begin{center}

\vspace{2mm}

{\Large \bf Schwarzschild black branes from unstable D-branes}

\vspace{3mm}

Oren Bergman${}^1$ and  Gilad Lifschytz${}^2$

\vspace{1mm}

${}^1${\small \sl Department of Physics} \\
{\small \sl Technion, Israel Institute of Technology} \\
{\small \sl Haifa 32000, Israel}\\
{\small \tt bergman@physics.technion.ac.il}
\vspace{1mm}

${}^2${\small \sl Department of Mathematics and Physics and CCMSC} \\
{\small \sl University of Haifa at Oranim, Tivon 36006, Israel} \\
{\small \tt giladl@research.haifa.ac.il}

\end{center}

\vskip 0.3 cm

\noindent
We study systems with a large number of meta-stable D$p$-branes,
and show that they describe Schwarzschild and Schwarzschild-like black
branes, generalizing the results of Danielsson, Guijosa and Kruczenski
\cite{dgk}. 
The systems are considered in both the open and closed string pictures.
We identify the horizon size and its relation to the physics of open 
and closed strings. From the closed string perspective the region 
inside the horizon is where the effects of massive
closed strings become important.

\end{titlepage}

\section{Introduction}
Low energy closed string theory possesses classical solutions describing many
types of black branes, both neutral and charged under NSNS or
RR fields. From among all these solutions the ones carrying RR charges
have received the most attention in the last eight years, starting with
\cite{sv}. It was shown
that D-branes provide a good description of states that evolve into
black holes as the number of branes is increased. This lead to
the AdS/CFT conjecture \cite{malda} and its generalizations \cite{imsy},
which provide a complete description of certain space-times, and in particular
black holes in them, by a gauge theory.
However this line of thought does not seem to work for uncharged
black holes, which do not have a D-brane description. 
The usual view on Schwarzschild-like black holes is that a massive string
state will become a black hole at a high enough mass for fixed coupling. 
The problem is
that the Schwarzschild-like black hole entropy cannot be accounted for by the
entropy of a string. Indeed, a black hole with a string size horizon
has an entropy $S \sim l_{s} M$ which is the maximal entropy that a closed
string can have. For a larger black hole the entropy has a different
dependence on the mass, and is greater than the entropy of any elementary string.
The entropy of a Schwarzschild black $p$-brane in
$D=p+d+3$ dimensions is given by
\begin{equation}
S=2\pi \omega_{d+1}^{-1/d}\left(\frac{2}{d+1}\right)^{(d+1)/d}
\kappa^{2/d}
V_{p}^{-1/d}M^{(d+1)/d} \;,
\end{equation}
where $\omega_{d+1}=2\pi^{1+d/2}/\Gamma(1+d/2)$ is the volume of a unit $(d+1)$-sphere.
This lead the authors of \cite{hkrs} to  propose that the density of states
of the string changes at large density to account for the larger entropy.
Another way of forming the black hole is to use a large collection of
unstable branes (or branes and anti-branes). One would expect
that tachyon condensation will generically produce an uncharged black hole.
An example of this is the known solution of Penrose of two colliding
plane waves which form a black hole. This is dual to a brane
anti-brane system.

Recent studies of the rolling tachyon background \cite{sen1,sen2} 
have produced new insights into this issue. 
In particular two results are important. First, it was shown
that the rolling tachyon produces mostly
massive closed strings which stay very close to the location of the original 
brane \cite{llm,gir}. This provides a dual closed string description
of tachyon matter.
Second, it was argued that certain closed string states
are just on the verge of forming unstable branes which can come in a
variety of dimensions \cite{gir}.

Consider some massive closed string, and fix the string coupling
to some constant value.
We can add more energy to the string until its Schwarzschild radius gets
to be of the order of the string scale. Now at this point adding more
energy will not increase the entropy as much as forming a Schwarzschild
black hole, which means that the probability of this string to stay a
massive close string state is very small. But the increased entropy
seems to indicate that there are new degrees of freedom available for
the system. What could they be? The lessons from the rolling tachyon
indicate that these degrees of freedom may just be open strings forming on
unstable D-branes (or on $D-\bar{D}$ systems, depending on the
dimension needed). At some point in the $g_{s}, M$ plane it becomes more
favorable to invest a certain energy in forming the branes, and thus
opening up a larger space of degrees of freedom.

Recently it was shown by Danielsson,
Guijosa and Kruczenski (DGK) \cite{dgk} that brane-anti-brane systems
based on D3, M2 and M5-branes
can describe the thermodynamics of Schwarzschild black holes in certain
dimensions. This was generalized to rotating 3-branes in \cite{ghm}.

In this note we will generalize and extend these results,
and discuss the emerging picture.

Note added: while this paper was completed a paper \cite{sp}
with which there is some overlap appeared.

\section{Open string picture}
We would like to consider how the unstable branes can be used to
describe the Schwarzschild black holes in various dimensions.
If we take a collection of $N_{0}$ unstable branes then they will tend to
decay. Their decay produces, in one view, very massive closed
strings. At some point, if the density of the closed strings is large
enough, the probability to emit a closed string may equal the probability 
to absorb one,
and the system can reach a meta-stable configuration (note that this
cannot happen if the decay were predominantly into massless closed strings).
In the dual view pairs of open strings are created, and here again if
the density of the open strings is large enough then the probability of
creation can become the same as the probability of open strings to
annihilate back to form the brane. It is the latter, open string, view that we will
consider first. Since the specific heat will turn out to be negative
it is better to work in the microcanonical ensemble. We will consider
the situation in which some of the $N_0$ branes have decayed and have
produced open strings on the rest of the $N<N_0$ branes.\footnote{Another
possibility is that all the branes decay a bit. It was suggested in \cite{dgk} that after 
stabilization there will still be massless open string excitations. One possibility is 
that since the system is stabilized there are moduli to move the center of mass, which 
should be just the zero-modes of some massless field on the branes. It is however not
clear how this is consistent with observations that the rolling of the tachyon gives mass 
to all open string excitations \cite{strominger}.  This may be taken into account by 
changing the 
tension of the branes to be a parameter as well. We will not consider this possibility here.}
As we will see, if $N_{0}$ is large
enough the resulting open strings will tend to be mainly in their
massless sector. 
The energy and entropy of the remaining meta-stable branes are given by
\begin{eqnarray}
M=N\tau_{p}V+a V N^{c}T^{\alpha+1}\\
S=\frac{\alpha +1}{\alpha} a V N^{c} T^{\alpha} \;,
\label{ms}
\end{eqnarray}
where $\tau_{p}$ is the unstable brane tension and  $V$ is its volume.
Note that $T$ is the temperature on the unstable brane.
Since we want to work in the microcanonical ensemble we will write
\begin{equation}
S=\frac{\alpha +1}{\alpha} a^{\frac{1}{\alpha +1}}V^{\frac{1}{\alpha
+1}}N^{\frac{c}{\alpha +1}}(M-N\tau_{p}V)^{\frac{\alpha}{\alpha +1}} \;.
\label{microentropy}
\end{equation}
To find the meta-stable configuration we need to maximize
the entropy with respect to $N$. This turns out to give
\begin{equation}
N=\frac{cM}{(\alpha+c)\tau_{p}V}
\label{nm}
\end{equation}
and thus
\begin{equation}
E=\frac{\alpha}{\alpha +c}M,
\end{equation}
where $E$ is the energy stored in open string excitations.
We now plug this back into the entropy expression to get
\begin{equation} 
S=(\alpha+1)a^{\frac{1}{\alpha +1}}\alpha^{\frac{1}{\alpha
+1}}(c+\alpha)^{-\frac{c+\alpha}{\alpha +1}}c^{\frac{c}{\alpha
+1}}\tau_{p}^{-\frac{c}{\alpha +1}}V^{\frac{1-c}{\alpha
+1}}M^{\frac{c+\alpha}{\alpha +1}} \;.
\label{entropy}
\end{equation}
This does not yet look like the entropy of the Schwarzschild black
brane.
What values of $c$, $a$ and $\alpha$ should we use?
The branes are now meta-stable and we are looking in a regime
where
$N$ is large, so this will not be a perturbative regime. We will now
make an assumption which we will justify by the
results we get. We will assume that the thermodynamics of the 
meta-stable $p$-brane in the large $N$ regime is the same as the thermodynamics of
the stable (i.e. BPS) brane in the large $N$ regime.

The thermodynamics of the stable branes is given by the formula \cite{imsy}
\begin{equation}
S_{stable}=h_{p}V
g_{YM}^{2\left(\frac{p-3}{5-p}\right)}N^{\frac{7-p}{5-p}}T^{\frac{9-p}{5-p}}\;,
\end{equation}
where $h_{p}$ is a known constant.
Putting this all into the formula in equation (\ref{entropy})
we get (dropping numerical constants and remembering that $g_{ym}^{2}
\sim g_{s}l_{s}^{p-3}$ and $\tau_{p} \sim (g_{s}l_{s}^{p+1})^{-1}$)
\begin{equation}
S \sim g_{s}^{\frac{2}{7-p}}l_{s}^{\frac{8}{7-p}} V^{\frac{-1}{7-p}}
M^{\frac{8-p}{7-p}}\;,
\end{equation}
which is exactly the right dependence on these parameters for
Schwarzschild black branes, although only up to a numerical factor of order one.
The negative specific heat of the Schwarzschild black brane is understood as the 
statement that
extra energy added to the system prefers to go into creating another
unstable brane, thereby increasing the number of degrees of freedom, and
lowering the temperature of the original excitations \cite{dgk}.

However there is a problem with this interpretation. The unstable
branes (or the brane anti-brane system) have a dilaton charge, while
the Schwarzschild black branes do not. We will
discuss this issue in the last section.

Since the temperature $T$ was taken to be the temperature of the gas
on the brane following the usual field theory relationship between
energy of the gas and its entropy, it is not clear that this will also be
the temperature of the black hole. So let us compute this. From equation (\ref{ms}) we find
\begin{equation}
\frac{1}{T}=\frac{\alpha}{\alpha+1}\frac{S}{M-N\tau_{p}V}
\end{equation}
Using the solution for $N$ (\ref{nm}) and the value of $\alpha$ one finds
\begin{equation}
\frac{1}{T}=\frac{8-p}{7-p}\frac{S}{M}\sim (g_{s}N)^{\frac{1}{7-p}}l_{s}\;,
\label{temp}
\end{equation}
which is exactly the black hole temperature. Notice that since we are
in the regime where $g_{s}N$ is large, the temperature is very low,
thus justifying the approximation of taking only the massless open
strings into account. At small horizon sizes of order the string length
the massive open strings have an important contribution to the
physics.

\subsection{More general brane configurations}

We can consider more general brane configurations
that arise in the low energy supergravity theory. 
We limit ourselves to the simplest
case described by having a gravity theory in D-dimensions with a
dilaton and a single $(p+1)$-form potential:
\begin{equation}
S=-\frac{1}{2\kappa^{2}}\int d^{D}x\sqrt{g}\left(R-\frac{1}{2}(\partial
\phi)^{2}-\frac{1}{2(d+1)!}
e^{a\phi}|F_{d+1}|^2\right)
\end{equation}
Black brane solutions for this action were analyzed in
\cite{kt}, and their near-extremal thermodynamics was extracted.
The entropy of the near-extremal $p$-brane solutions and the mass of
the extremal solutions are given by ($D=p+d+3$)
\begin{eqnarray}
S(E)= b
L^{p(1-\lambda)}\kappa^{\frac{2}{d}-\frac{n}{2}}q_{p}^{\frac{n}{2}}E^{\lambda}\\
b=2^{\frac{1}{d}-\frac{n}{4}+2}\pi
\omega_{d+1}^{-\frac{1}{d}}d^{-\frac{d+1}{d}}\lambda^{-\lambda}n^{-\frac{n}{4}}\\
M_{p,0}=\frac{q_{p}\sqrt{n}}{\sqrt{2}\kappa}L^{p} \;,
\end{eqnarray}
where $E$ is the energy above extremality, $L^p$ is the volume
of the $p$-dimensional torus which the brane wraps and
\begin{equation}
 \lambda=\frac{d+1}{d}-\frac{n}{2} \;,\;
 n=4\left[a^2+\frac{2d(p+1)}{D-2}\right]^{-1} \;.
\end{equation}
For the extremal solution to be supersymmetric $n$ has to be an
integer, and it then represents the number of different brane types in the
solution. Note that if we take $D=10$ and $n=1$ this reduces to the
example of the previous section.  
If we now assume the existence of unstable branes which become meta-stable
as before (or consider systems with branes and anti-branes), with
the thermodynamical properties as above, we can again consider systems
in which the number of these brane-configurations can vary, i.e. we write
\begin{eqnarray}
M=f M_{p,0}+E\\
S=S(E)=hS(M-M_{p,0}) \;.
\end{eqnarray}
The coefficient $f$ has to do with the ratio of tension of the unstable
branes to the stable ones,
and $h$ is some numerical coefficient that may be present.
We look for the number of branes (proportional to $q_{p}$) that
maximizes $S$ for a given $M$. We find
\begin{eqnarray}
q_{p}=\frac{\sqrt{2}\kappa}{f L^{p}}\frac{\sqrt{n}}{2\lambda +n}M\\
E=\frac{2\lambda}{2\lambda +n}M \;,
\end{eqnarray}
and,
\begin{equation}
S=hf^{-\frac{n}{2}}\pi
(d+1)^{-\frac{d+1}{d}}2^{\frac{1}{d}-\frac{n}{2}+2}\omega_{d+1}^{-\frac{1}{d}}
L^{p(1-\frac{d+1}{d})}\kappa^{\frac{2}{d}}M^{\frac{d+1}{d}} \;,
\end{equation}
which is a factor $h(2f)^{-\frac{n}{2}}$ times the Schwarzschild 
black brane entropy. 

With this set of branes at hand we can look for non-dilatonic
solutions to avoid the issue of dilaton charge. This means putting
$a=0$.
At least for supersymmetric configurations where $n$ is an integer,
there are only a few possibilities. The M-branes in eleven dimensions
and the D3-brane in ten dimensions were considered in
\cite{dgk}. Other possibilities are the self dual string in six
dimensions ($n=2$, $p=1$, $d=2$), which gives  the Schwarzschild black string
in six dimensions, and ($n=3$, $p=1$, $d=1$), which is a Schwarzschild black string 
in five dimensions.
Two other non-dilatonic examples, the black hole in four and
five dimensions, have $\lambda=0$, and
will be discussed at the end of the section.

We see that there is a large class of examples of meta-stable
brane configurations that provide a microscopic model for Schwarzschild
and Schwarzschild-like black branes.

\subsection{Quasi-Particle description}

We saw that the black brane is described by the thermodynamics of the
meta-stable branes, which we assumed was similar to that of the BPS
branes at large $N$. Now the thermodynamics of the BPS branes was
shown to have a simple quasi-particle picture \cite{ikll,ikll1} (at
least for the simple $p$-branes in ten dimensions).
This picture accounts in a simple way for the thermodynamical relation,
and more importantly for the relation between area and entropy and
between horizon size entropy and temperature \cite{ikll2}.

It was shown that the identification of the number of quasi-particles with
the entropy, together with the decay rates of the quasi-particles,
explains the equality of the area in Planck units and the entropy.
Since the meta-stable branes follow the same thermodynamical
relationship it is reasonable to assume that this description applies
to the Schwarzschild black brane case as well, as advocated in \cite{ikll2}. 
Since some of
the mass of the black hole is stored in the brane tension, only
a fraction $\frac{9-p}{16-2p}$ of the mass $M$ of the black hole is
carried by the quasi-particles' energy.

 From the point of view of the theory on the brane the horizon size is
given by a regulated expression for the size of the state
\cite{susskind,ikll}.
One takes the scalar fields $X_{i}$ which describe the transverse fluctuations of
the brane, and regularizes  $\mbox{Tr}\langle X^2\rangle$  by integrating out 
all the very massive
states. As shown in \cite{ikll} the field theory has a double peak
distribution which makes this procedure well defined, by including in
the trace just the first peak with frequency of order the
temperature. 
Since we assume
that the thermodynamics of the unstable branes is very similar to the
thermodynamics of the BPS branes, and since the horizon size is
related to the spectral density, one can expect that the expression for
the size of the state stays the same for the unstable branes.
The horizon size for the thermal BPS branes in the low
temperature regime is given by 
\begin{equation}
(g_{s}Nl_{s}^{p-3})^{1/2} T\sim U_{0}^{\frac{5-p}{2}}=
(\frac{r_{0}}{l_{s}^{2}})^{\frac{5-p}{2}}\;.
\end{equation}
Using equation (\ref{temp}) we then get
\begin{equation}
r_{0} \sim (g_s N)^{\frac{1}{7-p}}l_{s}\sim \frac{1}{T}\;,
\label{size}
\end{equation}
which is the correct relation for a Schwarzschild black brane. 
It is remarkable that the relationship between horizon size and
temperature for the Schwarzschild black brane follows from the properties of
the low temperature field theory on the brane.

The double peaked distribution for the spectral density of
the scalar also predicts the  relation \cite{ikll}
\begin{equation}
U_{0}^{2}\sim \frac{g_{s}N S}{V N^2 T} \;.
\end{equation}
This has been shown to be correct for the near horizon geometries of
\cite{imsy}.
Since the spectral densities should be similar in the meta-stable case,
this should be true here as well.
Since $VN/g_{s}\sim M$ we see that this relation is again just
\begin{equation}
r_{0}\sim 1/T.
\end{equation}
One might be worried that we are not in the regime where the
thermodynamical expressions are valid (i.e. where in the dual supergravity
the curvature is small at the horizon). However this is not the case,
the validity of the thermodynamical expressions we are using is in the
region \cite{imsy}
\begin{eqnarray}
r_{0} \ll (g_{s}N)^{1/(3-p)}l_{s}, \ \ p <3\\
r_{0} \gg (g_{s}N)^{1/(3-p)}l_{s} , \ \ p>3
\end{eqnarray}
which are satisfied for the value from equation (\ref{size}).

\subsubsection{General case}

All this was true for the simple $p$-branes in ten dimensions. For the
more general solution there is a similar relationship between
temperature and horizon size \cite{kt}
\begin{eqnarray}
E \sim r_{h}^{d}\\
T^{-1} \sim q_{p}^{n/2}E^{\lambda-1} \;,
\end{eqnarray}
where $r_{h}$ is the horizon radius.
This implies upon inserting $q_{p} \sim E$ that
\begin{equation}
T^{-1}\sim E^{1/d} \sim r_{h}\;,
\end{equation}
which is the correct relationship for a Schwarzschild black brane.

\subsection{$\lambda=0$ case}

Two other examples of non-dilatonic black holes are
($n=3$, $p=0$, $d=2$), which is a black hole in five
dimensions, 
and  ($n=4$, $p=0$, $d=1$),
which is a black hole in four dimensions.
For these cases $\lambda = 0$ and a more careful analysis is needed.

We start with the solutions from \cite{kt},
\begin{equation}
M=\frac{\sqrt{2}q_{p}L^p}{\kappa
\sqrt{n}d}\frac{d+1+nd\sinh^{2}\gamma}{\sinh 2\gamma} \;.
\end{equation}
Expanding near extremality (large $\gamma$) to second order we find
\begin{equation}
E = M - M_{p,0} = M_{p,0}\left[{4\over n}\lambda e^{-2\gamma} + 2 e^{-4\gamma}\right] \;.
\end{equation}
For $\lambda = 0$ we therefore get 
\begin{equation}
E=2M_{p,0}e^{-4\gamma} \;.
\end{equation}
In this case the entropy near extremality is given by
\begin{equation}
S=\omega^{-\frac{1}{d}}_{d+1}L^{p}\left(\frac{2\sqrt{2}\kappa
q_{p}}{\sqrt{n}d}\right)^{\frac{d+1}{d}}
\left(1+n\sqrt{\frac{E}{2M_{p,0}}}\right) \;,
\end{equation}
and the temperature by
\begin{equation}
\frac{1}{T}=\omega^{-\frac{1}{d}}_{d+1}L^{p}\left(\frac{2\sqrt{2}\kappa
q_{p}}{\sqrt{n}d}\right)^{\frac{d+1}{d}}\frac{n}{2\sqrt{M_{p,0}}}E^{-\frac{1}{2}}\;.
\end{equation}

We can now do the same calculation as before, fixing the total mass $M$ and
varying the number of branes (proportional to $q_{p}$ or $M_{p,0}$)
to maximize the entropy of the unstable brane (or brane-anti-brane
system). As before let us assume that the tension of the unstable brane
is given by $\tau =f \tau_{p}$ where $f$ is
some unknown constant.
We find that the entropy is maximized for
\begin{equation}
M_{p,0}=\frac{\alpha}{f} M\;,
\end{equation}
where $\alpha=\frac{13 +\sqrt{7}}{18}$ in the $n=4$ case, and
$\alpha=\frac{7 +\sqrt{5}}{11}$ for the $n=3$ case.
Plugging this into the expressions for the entropy and temperature we get
\begin{eqnarray}
S \sim M^{\frac{d+1}{d}}\\
T \sim M^{-\frac{1}{d}}.
\end{eqnarray}
Thus we have a description of the Schwarzschild black hole in four
and five dimensions.

\section{Closed string picture}

Now let us see how this is viewed from the closed string perspective.
The AdS/CFT and its generalizations have
taught us that the density of states of the closed string changes in
the presence of BPS branes. The procedure of zooming in on the near horizon 
geometry is just zooming in on the properties of the closed string
theory near the branes. The AdS/CFT claim is that the gauge theory is
dual to the closed string theory, i.e. that the number of states in the
closed string theory is given by the number of gauge invariant
configurations in the gauge theory. This number has the naive Hagedorn
behavior only at low enough energy, above which the density of states
of the closed string has a different dependence on the parameters
(see for example \cite{Banks}). 
If we assume that the density of states of the closed string near the 
unstable branes (after meta-stabilization) is similar to the density of states near 
BPS branes, then the results of the previous section can be interpreted
in terms of closed strings.

We thus view equation (\ref{microentropy}) as giving the density of
states
of the closed string at large levels in the presence of a large number $N$
of $p$-dimensional boundaries
\begin{equation}
\ln \Omega(E)\sim V^{\frac{5-p}{14-2p}}g_{s}^{\frac{p-3}{14-2p}}N^{1/2}E^{\frac{9-p}{14-2p}}\;,
\end{equation}
where $E$ is the energy of the closed string.
For the more general configurations one has 
\begin{equation}
\ln \Omega(E)\sim V^{1-\lambda}\kappa^{\frac{2}{d}-\frac{n}{2}}q_{p}^{n/2}E^{\lambda}\;.
\end{equation}
%Note that that the energy appears with a power less than or equal to one (for
%$p\leq 5$),
%although in the final answer for the black hole the power of the mass
%of the black hole is larger than one, which is due to the power of
%$N$ (or $q_{p}$).

While the unstable brane wants to decay into
massive closed strings, the massive closed strings produced do not move
away from the brane appreciably \cite{llm,gir}, thus enabling the
decay product to recombine to form the brane. At some point the decay
then stops and the brane becomes meta-stable by virtue of
the gas of massive closed strings. It is only meta-stable since
massless closed strings can be produced by the decay of the massive
closed strings and then escape as Hawking radiation.

\subsection{The horizon}
The appearance of a horizon is one of the puzzles of the black hole
issue. While this is well understood in the context of the low energy
supergravity, it is not clear what this means in the full string
theory.
In the GR context the horizon is locally not a special place.
As far as we know closed  string theory is only defined through
its S-matrix elements, which means that the asymptotic coordinate
system has a preferred status, unlike in classical GR where one
can change to different coordinates without any issue. This is also
amplified in the AdS/CFT context in which the asymptotic coordinates,
which break down near the horizon according to classical GR, are tied to the gauge
theory Hamiltonian one is using. In the AdS/CFT case it was argued
that the horizon position can be described as the locus of points
where new massless degrees of freedom (that can become tachyonic) appear when a D-brane probe
approaches the horizon \cite{kl,kl1}. This description is through the gauge theory
variables, but one can ask what is the dual closed string view.

In other examples where open string tachyons appear, a dual closed string
description states that the whole tower of massive closed
string modes becomes important, and dominates over the contribution of the
massless closed string modes. The appearance of a new massless open string
state away from the
origin of moduli space signals the breakdown of supergravity and the
dominance of massive closed string states. This is consistent with the idea
that the black hole states are comprised of many highly excited
massive closed strings. This is also the case when one considers the
unstable brane description. 
The horizon is where the low energy supergravity description breaks
down due to massive closed string effects, so the horizon is an
$\alpha'$ effect (or $g_{s}N$ effect). This is also in line with the 
analysis of \cite{lm,mat}, where it was shown that in certain cases the
horizon
indicates large fluctuations in the metric, and the breakdown of
uniqueness of the metric. It is also tempting 
to identify the singularity with the existence of the boundaries
(branes), being sources in closed string theory rather than on shell 
states.\footnote{This does not clarify what meaning, if any, 
the metric inside the horizon has.}

\section{Dilatonic Schwarzschild branes}

We saw that based on all charged branes one can get an object with the
correct dependence to be a Schwarzschild black brane. However as
mentioned above we expect some non trivial dilaton when the systems
are based on dilatonic branes. Now it may be possible that somehow the
dilaton is screened, but this seems unlikely. There are however more
general solutions to the Einstein-dilaton equations that have
a Schwarzschild-like dependence of the entropy on the energy, and
that satisfy the relation $T^{-1} \sim r_{H}$ \cite{zz} (see also \cite{bmo}).
The uncharged solutions are labeled by three parameters ($c_{1}, c_{3},r_{0})$.
However the requirement that the temperature be finite gives a relationship between
$c_{1}$ and $c_{3}$,
restricting them to a two parameter family \cite{oy}. 
There are some indications that the parameter $c_{1}=0$ (if $c_{3}=0$)
 corresponds to the tachyon at the top of the potential \cite{bmo}, although 
it is unclear if this 
is true also when $c_{3} \neq 0$.
 Confusing the picture a bit more is that the condition for finite temperature when 
$c_{1}=0$ cannot be satisfied unless $p=0$ or $p=3$, but can be satisfied for 
$c_{1} \neq 0$.
So there are at least two possibilities. One is that the top of the tachyon potential is
 unstable for the other values of $p$ and some rolling of the tachyon occurs before
 stabilization, and the other is 
that the relationship between the parameters and the value of the tachyon is more 
complicated. In any case 
it seems likely
that
the systems based on dilatonic branes describe these kinds of black branes.
A better understanding of this is clearly needed.

\section{Conclusion and Speculations}

We saw that near extremal charged black branes carry the information
needed to describe Schwarzschild-like black branes. The generality of
this result may suggest some universality in the thermodynamics of the
corresponding field theories and in the density of states of closed
strings near boundaries. 
These results suggest that the techniques used to analyze near-extremal charged 
black branes can
be used to study a larger class of systems. In particular this suggests
that Schwarzschild black branes (and possibly charged black branes far from
extremality)
are described by a quasi-particle picture similar to the one
describing near extremal black branes. The horizon seems to be the region where
massive closed string effects are important, suggesting that at least
from this point of view the horizon is a special place.
If Schwarzschild black branes can be described by open strings on meta-stable
branes this may suggest an extension of the AdS/CFT correspondence
to an open-closed duality in which a purely open string theory on any
brane is dual to a closed string theory in the background of that brane.
This would be a stronger version of the open-closed duality proposed
recently by Sen \cite{Sen}.

\section*{Acknowledgments}
We thank O.~Aharony and S.~Mathur for useful discussions.
The work of O.B. is supported in part by the
Israel Science Foundation under grant no.~101/01-1.
The work of G.L. is supported in part by the
US-Israel Binational Science Foundation grant no.~2000359.

%%%%%%%%%%%%%%%%%%%%%%%%%%%%%%%%%%%%%%%%%%%%%%%%%%%%%%%%%%%%%%%%%%%%%%%%%%%%%%%%

\end{document}